\documentclass[copyright,creativecommons,noderivs]{eptcs}
\providecommand{\event}{GandALF 2011}
\usepackage[latin1]{inputenc}
\usepackage{amsmath}
\usepackage{amssymb}
\usepackage{underscore}
\usepackage{amsthm}
\usepackage{tikz}
\usetikzlibrary{automata,patterns}
\newtheorem{defi}{Definition}
\newtheorem{theo}[defi]{Theorem}
\newtheorem{lem}[defi]{Lemma}
\newtheorem{corol}[defi]{Corollary}

\newcommand{\NN}{\mathbb{N}}

\newcommand{\AP}{\mathit{AP}}
\newcommand{\newterm}{\emph}

\title{Reactive Safety\thanks{This work was supported by the German Research Foundation (DFG) within the program ``Performance Guarantees for Computer Systems''
and the Transregional Collaborative Research Center ``Automatic Verification and Analysis of Complex Systems'' (SFB/TR 14 AVACS).}}
\author{R\"udiger Ehlers $\quad \quad$ Bernd Finkbeiner \\ \institute{
Reactive Systems Group\\Saarland University\\66123 Saarbr\"ucken, Germany}\email{\{ehlers,finkbeiner\}@cs.uni-saarland.de}}
\def\titlerunning{Reactive Safety}
\def\authorrunning{R. Ehlers \& B. Finkbeiner}
\begin{document}
\maketitle

\begin{abstract}
The distinction between safety and liveness properties is a
fundamental classification with immediate implications on the
feasibility and complexity of various monitoring, model checking, and
synthesis problems. In this paper, we revisit the notion of safety for
reactive systems, i.e., for systems whose behavior is characterized by
the interplay of uncontrolled environment inputs and controlled system
outputs. We show that reactive safety is a strictly larger class of
properties than standard safety. We provide algorithms for checking if
a property, given as a temporal formula or as a word or tree
automaton, is a reactive safety property and for translating such
properties into safety automata. Based on this construction, the
standard verification and synthesis algorithms for safety properties
immediately extend to the larger class of reactive safety.
\end{abstract}

\section{Introduction}
\label{sec:introduction}

The question whether a certain specified property,
given for example as a formula of a temporal logic, belongs to the class of \emph{safety
properties}, is of universal interest in verification, synthesis, and
monitoring.
Typically, it is much easier to reason about safety properties than about general temporal properties.
In deductive verification, safety properties are typically proven by
induction on the transition relation, while liveness properties
require a ranking function that maps the states into a well-founded
domain. In model checking, checking a safety property corresponds to
simple reachability, liveness to the more complicated nested
reachability.
In synthesis, deriving a system that satisfies a
safety property involves solving safety/reachability
games, which is simpler and typically more scalable than
solving games with more general winning conditions such as Muller or
parity.
Perhaps most significantly, in runtime analysis, safety
properties can be checked with a runtime monitor, while one can
never conclusively determine that a liveness property has been
violated after observing only a finite trace. 

We will refer to the standard definition of safety~\cite{Lamport/77/proving,DBLP:journals/ipl/AlpernS85} as \emph{linear-time safety},
because it is based on the linear-time semantics, where the system and the specification each define a set of infinite words over an alphabet of observations.
A language of infinite words is
a linear-time safety property iff for every word $w$ that violates $P$
(i.e., $w \not\in P$), there exists a \emph{finite prefix} $w'$ of
$w$ such that $w'$ also violates $P$, i.e., for all infinite
extensions $w''$ of $w'$ it holds that $w'' \not\in P$. In this paper, we show that
the class of safety properties can be significantly extended if, rather than considering
words over a single alphabet of observations, one explicitly distinguishes between the inputs and the outputs of a reactive system.

We introduce our new notion of \emph{reactive safety} by way of an example. Let us
use linear-time temporal logic (LTL) to specify a simple coffee
machine with two input bits $c$ (the coffee button)
and $e$ (emergency shutdown), and two outputs $b$ (brewing coffee) and
$f$ (emitting a failure signal).  We specify that whenever the user
presses the coffee button, brewing must eventually start or a failure
must be signaled immediately. As an LTL formula, this
property can be expressed as follows:\footnote{For this example, we
  assume that in every clock cycle, the system first generates the
  output and then reads its input.}
\begin{equation}
\psi_1 = \mathsf{G}(c \rightarrow \mathsf{X} ( f \vee \mathsf{F} b)).
\end{equation}
Additionally, we require that whenever the emergency shutdown button is pressed, brewing stops immediately (i.e., when the system gives the next output) and permanently:
\begin{equation}
\psi_2 = \mathsf{G}(e \rightarrow \mathsf{X} \mathsf{G}(\neg b)).
\end{equation}

Clearly, $\psi_2$ is a linear-time safety property and $\psi_1 \wedge
\psi_2$ is not, because there is no bound on the number of steps until
the brewing starts after the coffee button was pressed. However, $\psi_1 \wedge \psi_2$ is a
reactive safety property: we can transform $\psi_1 \wedge \psi_2$ into
a linear-time safety property $\psi'_1 \wedge \psi_2$ that is
equivalent in the sense that any system with input $2^{\{c, e\}}$ and output
$2^{\{b, f\}}$ satisfies $\psi_1 \wedge \psi_2$ if and only if it satisfies
$\psi'_1 \wedge \psi_2$. For $\psi_1'$, the safety formula
$\mathsf{G}(c \rightarrow \mathsf{X} f)$ can be used. To see this, observe that $\psi_1$ specifies that whenever the
coffee machine does not immediately respond to a coffee request with a
failure message, it must eventually brew coffee regardless of the further
circumstances. However, if the user presses the emergency shutdown
button, the system cannot fulfill this task anymore without violating
$\psi_2$. Thus, the only possibility for the system to satisfy $\psi_1
\wedge \psi_2$ is to answer every request with an immediate failure
message. \sloppy

A natural semantic setting for reactive safety is that of branching time,
where we view the computation
of the system as a tree that branches according to the environment
actions and where each node is labeled with the system's response to a
particular sequence of environment actions. Reactive safety should, however,
not be confused with existing notions of safety for tree properties, which 
extend safety from linear time to branching time by 
referring to prefix trees
rather than prefix words:   Manolios and
Trefler~\cite{DBLP:conf/lics/ManoliosT01,DBLP:conf/podc/ManoliosT03}
define a \emph{universal safety} property as a set $P$ of infinite
trees such that for every tree $t$ that violates $P$, there exists a
finite prefix tree $t'$ of $t$ such that $t'$ also violates $P$, i.e.,
for all infinite extensions $t''$ of $t'$ it holds that $t'' \not\in
P$.
The price for referring to prefix trees is that the
algorithmic advantages of linear-time safety are lost. For example,
the branching-time property $\theta$ that states that the system's reaction to
environment action $0$ is different to its reaction to environment
action $1$ (formally, the set of binary trees where the label on the
$0$-child of the root is different from the label on the $1$-child) is
universally safe. However, it is impossible to construct a runtime
monitor for this property, because the monitor cannot follow two
branches at the same time.

The notion of reactive safety applies uniformly to words and trees.
Stated in terms of a tree language, a set of infinite trees is a reactive safety property iff
for every tree $t$ that violates $P$, there exists a finite path $w$
in $t$ such that any tree $t'$ \emph{that contains $w$} also
violates $P$, i.e., it holds that $t' \not\in P$. We call the node
that is reached by $w$ the \emph{violation starting node} of $P$.
Stated in terms of a word language, a set of infinite words $P$ is a reactive safety property iff
the set of trees whose paths are contained in $P$ (we call
this set the \emph{spread} of $P$) is a reactive safety property.

The class of reactive safety properties lies strictly between
linear-time and branching-time safety: every linear-time safety property is
also a reactive safety property, because the violating prefix
identifies a violation starting node; likewise, every reactive safety
property is also a universal safety property, because the path to the
violation starting node is also a finite subtree. As our examples
show, the inclusion is strict: the coffee machine specification $\psi_1
\wedge \psi_2$ is a reactive safety property but not a linear-time
safety property; the branching-time property $\theta$ is a universal
safety property but not a reactive safety property.

In fact, one can view reactive safety as the natural connection point
between linear-time and branching-time safety. As we show later in the
paper, reactive safety characterizes precisely the
class of tree properties whose satisfaction can be
checked by testing if all paths satisfy some linear-time
safety property. Hence, reactive safety captures as much of the
generality of branching-time safety as one can afford if one wishes
to retain the algorithmic advantages of linear-time safety:
All standard constructions for the verification and
synthesis of linear-time safety properties can still be applied for
reactive safety properties.

In the remainder of the paper, we present algorithms for checking if a
property, given as a temporal formula or as a word or tree
automaton, is a reactive safety property and for automatically
translating such properties into linear-time safety properties, expressed as safety automata.  
An immediate
application of the algorithms is \emph{specification debugging},
where the developer is warned if a property is a reactive safety
property but not a linear-time safety property.
There can be several reasons for such a situation.  On the one
hand, the specification might be erroneous, which should be detected
as early as possible in the development process. On the other hand, an
implicit equivalence, such as the one between $\psi_1 \wedge \psi_2$
and $\psi'_1 \wedge \psi_2$, may be an intended consequence of the
specification. For the developer, this case is also of interest as it
may be possible to reformulate the specification in a more direct and
more concise way; understanding the consequences of the
specification is also helpful for the subsequent design decisions.

A second major application of our algorithms is to \emph{extend}
verification, synthesis and monitoring methods for linear-time safety
to reactive safety.  If a specification is a reactive safety property
but not a linear-time safety property, we automatically construct a
safety automaton, which represents a linear-time safety property that
is equivalent in the sense that it has the same meaning on all systems
with the same interface (i.e., the same inputs and outputs). The
safety automaton can thus replace the original property for any
verification, synthesis or monitoring purpose.

\paragraph{Related work.}
The advantages of safety
properties in verification (cf. \cite{Manna+Pnueli/89/Completing}), synthesis
(cf. \cite{DBLP:conf/fmcad/SohailS09}) and runtime monitoring
(cf. \cite{DBLP:journals/fmsd/FinkbeinerS04}) are discussed in numerous
papers and textbooks. However, determining whether a given property is a  safety property is also useful independently of these applications. For instance, in \newterm{specification debugging}, unintended properties of manually written specifications are to be found. Two well-known techniques in this context are 
\newterm{vacuity
  checking}~\cite{DBLP:journals/fmsd/BeerBER01}, which searches for
inconsistencies and tautologies in the specification, and testing for \newterm{semantical safety in the
linear-time paradigm} \cite{DBLP:journals/fmsd/KupfermanV01}, where LTL
formulas that express linear-time safety properties but possibly contain
operators like \emph{until} or \emph{eventually} 
are identified.
Our example specification
$\psi_1 \wedge \psi_2$ is neither vacuous nor 
semantical safe in the
linear-time paradigm, but still deserves a warning, because it can be
stated equivalently as the linear-time safety property $\psi_1' \wedge
\psi_2$. Thus, identifying reactive safety properties can be seen as a refinement of these two techniques.

The game-like view onto the interactions between inputs and outputs, which distinguishes reactive safety from the standard linear-time safety, has been used previously in related works. For instance, linear-time properties and their respective reactive safety properties in our framework are connected by the concept of open implication that was introduced by Greimel, Bloem, Jobstmann and
Vardi~\cite{DBLP:conf/icalp/GreimelBJV08}. A linear-time property has an equivalent reactive safety property if and only if both properties openly imply each other. Pnueli, Zaks and Zuck \cite{DBLP:journals/entcs/PnueliZZ06} furthermore applied the game-based viewpoint in the field of \newterm{runtime verification} and solved the \emph{interface monitoring problem} of universal liveness properties. 

\section{Preliminaries}

We consider non-terminating systems that interact with their
environment over an infinite run. 
The interface between the system and
the environment is specified by a \emph{signature} $(I,O)$, where $I$ and $O$ are two disjoint sets of 
input and output signals, respectively. 
Each sequence of inputs results in a sequence of
outputs. We therefore formalize system runs as infinite words over $O
\times I$, and complete system behaviors as infinite $O$-labeled trees
that branch according to $I$. In this section, we give a quick summary
of the standard terminology for infinite words and trees. We
also describe linear-time temporal logic as an example logic for the
specification of reactive systems, and automata on infinite words and
trees, which provide the basic machinery for the constructions of the
paper. For a more detailed background on word and tree automata
in the context of reactive systems, the reader is referred to~\cite{DBLP:conf/birthday/Thomas08}. \\

\noindent
{\bf Words. } Given some finite alphabet $\Sigma$, we denote with $\Sigma^*$ and $\Sigma^\omega$  the sets of finite and infinite words over $\Sigma$, respectively. 
For a reactive system with signature $(I,O)$, we use infinite words in $(O \times I)^\omega$ to represent runs, and finite words in $(O \times I)^*$ to reason about the prefixes of such runs. A word $w = (y_0,t_0), (y_1, t_1) \ldots$, with $y_i \in O$ and $t_i \in I$ for every $i \in \NN$, describes a run of a reactive system in which $y_0$ is put out in the first computation cycle, then $t_0$ is read and $y_1$ is put out, and so forth. This definition corresponds to the notion of \newterm{Moore automata} \cite{Mueller2000}. 

A subset of $\Sigma^\omega$  is called a \emph{word language} or a \emph{word property}. We say that a word $w$ \emph{satisfies} a word property $P$ iff $w \in P$. Given some word $w = w_0 w_1 \ldots$, we denote by $w^i = w_i w_{i+1} \ldots$ the \emph{suffix} of $w$ starting in position $i$. \\

\noindent
{\bf Linear-time temporal logic. }
\emph{Linear-time temporal logic (LTL)} \cite{DBLP:conf/focs/Pnueli77} is a commonly used logic to express properties over runs of a system. Formulas in LTL are defined with respect to a set of atomic propositions $\AP$. For a reactive system with signature $(I,O)$, we assume that there exists a corresponding pair of sets of atomic propositions $(\AP_I, \AP_O)$ such that $I = 2^{\AP_I}$ and $O=2^{\AP_O}$.  We set $\AP = \AP_I \cup \AP_O$. The syntax of LTL is defined inductively as follows:
\begin{itemize}
 \item For all atomic propositions $x \in \AP$, $x$ is an LTL formula.
 \item Let $\phi_1$ and $\phi_2$ be LTL formulas. Then $\neg \phi_1$, $(\phi_1 \vee \phi_2)$, $(\phi_1 \wedge \phi_2)$, $\mathsf{X} \phi_1$, $\mathsf{F} \phi_1$, $\mathsf{G} \phi_1$, and $(\phi_1 \mathsf{U} \phi_2)$ are also valid LTL formula.
\end{itemize}
\looseness-1 The validity of an LTL formula $\phi$ over $\AP$ is defined inductively with respect to an infinite word $w = w_0 w_1 \ldots \in (2^{\AP})^\omega$. Let $\phi_1$ and $\phi_2$ be LTL formulas. We set:
\begin{itemize}
 \item $w \models p$ if and only if (iff) $p \in w_0$ for $p \in \AP$
 \item $w \models \neg \psi$ iff not $w \models \psi$
 \item $w \models (\phi_1 \vee \phi_2)$ iff $w \models \phi_1$ or $w \models \phi_2$
 \item $w \models (\phi_1 \wedge \phi_2)$ iff $w \models \phi_1$ and $w \models \phi_2$
 \item $w \models \mathsf{X} \phi_1$ iff $w^1 \models \phi_1$
 \item $w \models \mathsf{G} \phi_1$ iff for all $i \in \NN$, $w^i \models \phi_1$
 \item $w \models \mathsf{F} \phi_1$ iff there exists some $i \in \NN$ such that $w^i \models \phi_1$
 \item $w \models (\phi_1 \mathsf{U} \phi_2) $ iff there exists some $i \in \NN$ such that for all $0 \leq j < i$, $w^j \models \phi_1$ and $w^i \models \phi_2$
\end{itemize}

Given an LTL formula $\psi$ over $\AP$, the set of words satisfying the formula is a word language over $2^\AP$, denoted as $\mathcal{L}(\psi)$.\\

\noindent
{\bf Word automata. }
Like LTL formulas, word automata represent word languages. Formally, a (universal or nondeterministic) \emph{parity word automaton} is a tuple $\mathcal{A} = (Q,\Sigma,\delta,q_0,\alpha)$, where $Q$ is a finite set of states, $\Sigma$ the alphabet of $\mathcal{A}$, $\delta : Q \times \Sigma \rightarrow 2^Q$ the transition function of $\mathcal{A}$, $q_0 \in Q$ the initial state and $\alpha : Q \rightarrow \NN$ is the coloring function of $\mathcal{A}$. If $\alpha$ maps all states to 0 or 1, then $\mathcal A$ is called a
\emph{B\"uchi
  automaton}. If $\alpha$ maps all states to $0$, then $\mathcal{A}$ is called a \emph{safety automaton}. In this case, we omit $\alpha$ from the tuple. 

To determine if a given word $w = w_0 w_1 \ldots \in \Sigma^\omega$ is in the language of the word automaton $\mathcal A$ (we also say $w$ is \emph{accepted} by $\mathcal{A}$) we consider the runs of $\mathcal A$ on $w$. A \emph{run} on $w$ is a sequence $\pi = \pi_0 \pi_1 \ldots \in Q^\omega$ such that $\pi_0 = q_0$ and for all $i \in \NN$, $\pi_{i+1} \in \delta(\pi_i,w_i)$. We say that $\pi$ is an \emph{accepting run} if $\max(\inf(\pi))$ is even, where $\inf$ is the function that maps the sequence $\pi$ to the elements occurring infinitely often in it.

If $\mathcal{A}$ is a \newterm{nondeterministic automaton}, then $\mathcal{A}$ accepts the words for which there exists an accepting run. On the other hand, if $\mathcal{A}$ is a \newterm{universal} automaton, then $\mathcal{A}$ accepts those words for which all infinite runs for the word are accepting. We call a nondeterministic automaton where, for all $q \in Q$, $x \in \Sigma$, we have $|\delta(q,x)| \leq 1$, \newterm{deterministic}.

The connection between LTL and word automata is well-established in the literature. An LTL formula can be converted to an equivalent Büchi automaton of size exponential in the length of the LTL formula \cite{Vardi94reasoningabout}, where we define the size of an automaton to be $|\Sigma| \cdot |Q|$. \\

\noindent
{\bf Trees. } We use words to describe runs of a reactive system and trees to describe the overall behavior of a reactive system, i.e., its output for all possible sequences of inputs. Given finite sets $I$ and $O$, we define the set of \newterm{$O$-labeled $I$-trees} $O^\omega_I$ as all pairs $\langle T, \tau \rangle$ such that $T \subseteq I^*$ is a prefix-closed set and $\tau : T \rightarrow O$ is a function that labels each node of the tree with an element of $O$. 
We call $I$ the set of \newterm{directions} of the tree and $O$ its \newterm{set of labels}. Whenever clear from the context, we omit $I$ and $O$ and just call $\langle T, \tau \rangle$ a tree. 
We call a tree $\langle T, \tau\rangle$ for which $T = I^*$ holds, a \newterm{full tree}.
A \emph{tree property} or \emph{tree language} $\psi$ over $I/O$-trees is a subset of $O^\omega_I$. 
A tree $\langle I^*, \tau \rangle$ with $\tau : I^* \rightarrow O$ is a representation for a reactive system with signature $(I,O)$. The runs of the reactive system correspond to the paths through the tree, i.e., each run is a word $\pi = s_0 t_0 s_1 t_1 \ldots \in (O \times I)^\omega$ such that for every $n \in \NN_0$, $t_0 t_1 \ldots t_{n-1} \in T$ and $\tau(t_0 \ldots t_{n-1}) = s_n$.
We say that $\pi$ is \emph{maximal} if $\pi$ is infinite or for $\pi = s_0 t_0 s_1 t_1 \ldots s_n t_n$, for no $x \in I$, we have $t_0 \ldots t_{n} x \in T$.\\

\noindent
{\bf Tree automata. } We use tree automata to define properties of the overall behavior of a reactive system. A (nondeterministic or universal)
 \emph{parity tree automaton} is a tuple $\mathcal{A} =
(Q,I,O,\delta,q_0,\alpha)$ with a finite set of states $Q$, a
finite set of directions $I$, a finite set of labels $O$, a
transition relation $\delta \subseteq Q \times O \times (I
\rightarrow Q)$, and a coloring function $\alpha : Q \rightarrow
\NN$. We say that a tree automaton $\mathcal{A}$ is
deterministic if for each $q \in Q$ and $y \in O$, there exists
at most one element of the form $(q,y,f)$ for some $f \in (I
\rightarrow Q)$ in $\delta$.  As for word automata, we call $\mathcal{A}$ a \emph{safety
  automaton} if $\alpha$ maps all states to $0$ and a \emph{B\"uchi
  automaton} if $\alpha : Q\rightarrow \{0,1\}$.

Given an $O$-labeled $I$-tree $\langle T, \tau \rangle$, we say
that some $Q$-labeled $I$-tree $\langle T_r, \tau_r \rangle$ is a \emph{run
tree} of $\mathcal{A}$ and $\langle T, \tau \rangle$ if
$\tau_r(\epsilon) = q_0$ and for all $t \in T_r$, there exists some $f
\in (I \rightarrow Q)$ with $(\tau_r(t),\tau(t),f) \in \delta$
such that for all $x$ with $f(x)=q$ for some $q \in Q$, we have
$\tau_r(tx)=q$.  We say that $\langle T_r, \tau_r \rangle$ is an
\emph{accepting} run tree if $T_r = T$ and for all infinite paths $\pi = q_0
t_0 q_1 t_1 \ldots$ in $\langle T_r, \tau_r \rangle$, the
highest number occurring infinitely often in the sequence $\alpha(q_0)
\alpha(q_1) \ldots$ is even.  
For a nondeterministic parity tree automaton $\mathcal{A}$, we say that $\langle T, \tau \rangle$
satisfies $\mathcal{A}$ (and, equivalently, that $\langle T, \tau \rangle$ is accepted
by $\mathcal{A}$) if there exists an accepting run tree for $\langle T,
\tau \rangle$ and $\mathcal{A}$.  
A universal parity tree automaton $\mathcal{A}$ accepts a tree $\langle T, \tau \rangle$ if all full run trees for $\langle T, \tau \rangle$ are accepting. The language of $\mathcal{A}$,
written $\mathcal{L}(\mathcal{A})$, consists of all accepted trees.

For a state $q\in Q$ of a tree automaton $\mathcal{A} = (Q,I,O,\delta,q_0,\alpha)$,
we define the \emph{language of} $q \in Q$ as the language of the
automaton $\mathcal{A}' = (Q,I,O,\delta,q,\alpha)$. Likewise, the
language of a state $q \in Q$ in a word automaton $\mathcal{A} =
(Q,\Sigma,\delta,q_0,\alpha)$ is defined as the language of the
automaton $\mathcal{A}' = (Q,\Sigma,\delta,q,\alpha)$.

An automaton is called \emph{pruned} if it has no states with empty language.  We
define the \emph{size} of a tree automaton $\mathcal{A}$ as $|\mathcal{A}| = |Q| +
|\delta|$. We say that a tree or word property is a \newterm{regular property}
if it is the language of a parity tree or word
automaton, respectively. We say that $q_1 q_2 \ldots q_n \in Q^n$ for some $n \in
\NN$ is a cycle in a tree automaton $\mathcal{A}$ if $q_1 = q_n$ and for every $i \in
\{1, \ldots, n-1\}$ there exist $y \in O$ and $x \in I$ such that
$f(x)=q_{i+1}$ for some $f$ with $(q_i,y,f) \in \delta$.\\

\noindent
{\bf From word to tree properties. }
We often use word properties to describe the overall behavior of a
reactive system by requiring that every path of the tree satisfies the
word property: for example, a reactive system satisfies a
specification given as an LTL formula iff the LTL formula is satisfied
for all possible input sequences. To formalize the translation from
word to tree properties, we introduce a special spreading function.
The \newterm{spreading} $\mathcal{S}_{I/O}(\psi)$ of a word language $\psi \subseteq (O \times I)^\omega$ for a signature $(I,O)$ is defined as follows:
\begin{equation*}
\mathcal{S}_{I/O}(\psi) = \{ \langle I^*, \tau \rangle \mid \forall t = t_0 t_1 \ldots \in I^\omega: (\tau(\epsilon),t_0) (\tau(t_0),t_1) (\tau(t_0 t_1),t_2) \ldots \in \psi\}
\end{equation*}
It is straightforward to implement the spreading function as a construction that builds a tree automaton from a given deterministic parity word automata, such that the language of the tree automaton is the spreading of the the regular language represented by the word automaton.

\begin{defi}
\label{TFunction}
Given a deterministic parity word automaton $\mathcal{A} = (Q,\Sigma,\delta,q_0,\alpha)$ with $\Sigma = O \times I$, we define $\mathcal{T}_{I/O}(\mathcal{A}) = \mathcal{A}'$ for the deterministic tree automaton $\mathcal{A}' = (Q,I,O,\delta',q_0,\alpha)$ for which for all $q \in Q$, $x \in O$ and $f \in (I \rightarrow Q)$ we have $(q,x,f) \in \delta'$ if and only if for all $y \in I$, $f(y) = q'$ for some $q' \in Q$ if and only if $(q,(y,x),q') \in \delta$. 
\end{defi}

\bigskip 

\noindent
{\bf Linear-time and branching-time safety. }
Given a word language $\psi$ over some alphabet $\Sigma$, we say that $\psi$ is a \newterm{linear-time safety} property if for every $w = w_0 w_1 \ldots \in \Sigma^\omega$ such that $w \notin \psi$, there exists some $i \in \NN$ such that for all words $w' \in \Sigma$, $w_0 w_1 \ldots w_i w' \notin \psi$ \cite{DBLP:journals/ipl/AlpernS85}. The prefix $w_0 w_1 \ldots w_i$ is also called a \newterm{bad prefix word}. If $\psi$ is a regular property and also a safety property, then $\psi$ can also be represented as a safety word automaton.

Given some tree $\langle T, \tau \rangle$, we say that some tree $\langle T', \tau' \rangle$ is a finite prefix tree of $\langle T, \tau \rangle$ if $T' \subseteq T$, $T'$ is finite, and for all $t \in T'$, we have $\tau'(t) = \tau(t)$. A tree property $\psi$ over $I$/$O$-trees is  a \emph{universal safety property}~\cite{DBLP:conf/lics/ManoliosT01}
 if all trees, for which all finite prefix trees are the prefix of some tree in $\psi$, are also in $\psi$.

\section{Reactive Safety}
This section gives a formal definition of reactive safety. We start by considering general word and tree languages and will only later, in Section~\ref{regularreactivesafety}, focus on 
the special case of regular properties, as defined by automata or temporal logic formulas.
We show that the class of reactive safety properties lies strictly between linear-time safety and  universal safety. We also prove that reactive safety captures the largest class of properties whose satisfaction by a reactive system can be checked by testing whether all runs of the system satisfy some linear-time safety property.

Unlike standard linear-time safety, reactive safety distinguishes between inputs and outputs. We therefore parameterize
reactive safety with the signature of the reactive system and refer to
reactive safety with respect to signature $(I,O)$ as $I$/$O$-safety.
\begin{defi}
\label{def:IOSafety}
Let $I$ be a finite set of inputs, $O$ be a finite set of outputs, and let $\psi$ be a set of full
$O$-labeled $I$-trees.  We say that $\psi$ is a
\newterm{reactive safety property with respect to input $I$ and output $O$}, or short
an \emph{$I$/$O$-safety property}, if, for every $O$-labeled
$I$-tree $\langle T, \tau \rangle$ that is not contained
in $\psi$, there exists some node $t = t_0 \ldots t_k \in T$ (the
\emph{violation starting node}) such that all $I$/$O$-trees $\langle T',
\tau' \rangle$ for which $\tau(t_0 \ldots t_i) = \tau'(t_0 \ldots
t_i)$ holds for all $0 \leq i < k$, we have that $\langle T', \tau'
\rangle \notin \psi$. 
\end{defi}
Informally, reactive safety thus means that whenever a tree does not
satisfy the property, there exists some prefix path through the tree
such that at the end of the path, it is clear that there exists no
tree containing this prefix path such that the overall tree satisfies
the property. The notion of reactive safety extends to word properties:
A word property $\psi$ over the alphabet $O \times I$ is an $I$/$O$-safety property iff the spreading $\mathcal{S}_{I/O}(\psi)$ is an $I$/$O$-safety property. 
In the following, we omit $I$ and $O$ whenever clear from the context, and simply refer to
reactive safety. 

The difference between the definitions of linear-time  and reactive safety is subtle:
In the case of linear-time safety, a word is accepted iff it does not have a bad prefix; hence, on a
tree, every violating path must have a bad prefix. 
In the case of reactive safety, a tree is accepted iff it does not have a violation starting node: the difference thus is that for reactive safety, a single 
path to the violation starting node suffices for the entire tree, whereas for linear-time safety, every violating path needs to have a bad prefix.

We now compare reactive safety to linear-time and universal safety.
The following theorem shows that linear-time safety is a stronger requirement than reactive safety.
\begin{theo}
\label{theo:Sisreac}
Let $\psi$ be a linear-time safety word property over some alphabet $O \times I$. Then $\mathcal{S}_{I/O}(\psi)$ is a reactive safety property.
\end{theo}
\begin{proof}
Let $\langle I^*, \tau \rangle$ be a tree that is not contained in $\mathcal{S}_{I/O}(\psi)$. This means that there exists some path $t = t_0 t_1 \ldots \in I^\omega$ in the tree such that $w = (\tau(\epsilon),t_0) (\tau(t_0),t_1) (\tau(t_0 t_1),t_2) \ldots$ is not contained in the safety word property $\psi$. The definition of linear-time safety assures that then, there is also some prefix of length $k$ for some $k \in \NN$ and $t$ such that no word starting with $(\tau(\epsilon),t_0) (\tau(t_0),t_1) (\tau(t_0 t_1),t_2) \ldots (\tau(t_0 \ldots t_{k-1}), t_{k})$ is in $\psi$. In this case, we know that $t_0 \ldots t_k$ is a violation starting node in $\langle I^*, \tau \rangle$. Thus, all trees rejected by $\mathcal{S}_{I/O}(\psi)$ have a violation starting node, which makes $\mathcal{S}_{I/O}(\psi)$ a reactive safety property.
\end{proof}
The coffee machine example from the introduction shows that a reactive safety property is not necessarily also a linear-time property. 
Comparing reactive and universal safety, we immediately see that reactive safety is stronger than universal safety, because the path to the
violation starting node is also a finite subtree.
\begin{corol}
Every reactive safety property is also a universal safety property.
\end{corol}

The converse is not true. 
Formalizing the example property $\theta$ from the introduction, consider $I = \{0,1\}$, $O = \{0,1\}$ and the tree property $\theta = \{ \langle T, \tau \rangle \mid T = I^*, \tau(0) \neq \tau(1) \}$. This property is certainly universally safe, but not a reactive safety property, because it relates the labels along two paths; a violation can therefore not be blamed on a single violation starting node.

Reactive safety thus lies strictly between linear-time and universal safety. As discussed in the introduction, one can in fact view reactive safety as the natural connection point between linear-time and branching-time safety, because it represents
the largest class of properties whose satisfaction by a reactive system can be checked by testing whether all runs of the system are contained in some linear-time safety property.
This characterization of reactive safety is proven in the following theorem.
\begin{theo}
A tree property $\psi \subseteq O^\omega_I$ is an $I$/$O$-safety property iff there exists a word property $\psi' \subseteq (O \times I)^\omega$ such that precisely the trees in $\psi$ satisfy $\psi'$ along all of its paths.
\end{theo}
\begin{proof}
The ``if'' direction is implied by Theorem~\ref{theo:Sisreac}.
For the ``only if'' direction, we define $\psi'$ to contain all paths in trees in $\psi$ that do not contain a violation starting node. Then, $\psi'$ accepts the words needed for the claim. Surely, $\psi'$ is also a safety property as for every path not in $\psi'$, the path must contain a violation starting node, and every other path with the same prefix up to this node is also not in $\psi'$.
\end{proof}
We conclude this section by returning to the coffee machine example from the introduction. We specified the coffee machine with the two LTL formulas
\[
\psi_1 = \mathsf{G}(c \rightarrow \mathsf{X} ( f \vee \mathsf{F} b)) \qquad \mbox{and} \qquad
\psi_2 = \mathsf{G}(e \rightarrow \mathsf{X} \mathsf{G}(\neg b)).
\]
The conjunction $\psi_1 \wedge \psi_2$ is an $I/O$-safety property for the signature $(I = 2^{\{c,e\}}$, $O = 2^{\{b,f\}})$. To see this, consider a tree $\langle T, \tau \rangle$ that does not fulfill $\psi_1 \wedge \psi_2$ along all of its paths. Violation starting nodes are:
\begin{enumerate}
 \item the nodes that witness that $\psi_2$ has been violated along the path to the node, and
 \item the nodes $t = t_0 \ldots t_k$ for which $c \in t_{k-1}$, but $f \notin \tau(t_0 \ldots t_{k-1})$ and $e \in t_k$, as any such prefix path $(\tau(\epsilon),t_0) (\tau(t_0),t_1) \ldots (\tau(t_0 \ldots t_{k-1}),t_k)$ cannot be extended to an infinite path that satisfies $\psi_1 \wedge \psi_2$ (as explained in the introduction).
\end{enumerate}
It is not obvious to see that the set of trees satisfying $\psi_1 \wedge \psi_2$ is precisely the set of trees that do not have a violation starting node corresponding to one of the two node types above. In the next sections we will develop the necessary automata-theoretic machinery to answer this question. We will return to the example in Section~\ref{examplecontinued}.

\section{Regular Reactive Safety Properties}
\label{regularreactivesafety}

We now give an automata-theoretic characterization of the regular
$I$/$O$-safety tree properties.
Let $\psi$ be an $I$/$O$-safety property. In analogy to the
definition of tight automata for linear-time safety
languages~\cite{DBLP:journals/fmsd/KupfermanV01}, we call a
deterministic word automaton $\mathcal{A}$ \newterm{tight} for $\psi$
if $\mathcal{T}_{I/O}(\mathcal{A})$ accepts precisely the trees
$\psi$. In the following, we establish the fact that all regular reactive safety properties have regular tight languages, which immediately implies that the class of deterministic safety tree automata represents precisely the reactive safety languages.

The key step is to define a function $\mathcal{W}$, which converts a tree
automaton to a word automaton. Intuitively, $\mathcal{W}$ is the inverse operation to spreading a word automaton. The $\mathcal{W}$ function is the missing link in the characterization of the regular reactive safety properties -- we show that a property, represented as a (pruned) tree automaton $\mathcal{A}$ is $I$/$O$-safe if and only if we have $\mathcal{L}(\mathcal{A}) = \mathcal{L}(\mathcal{T}_{I/O}(\mathcal{W}(\mathcal{A})))$. 

We begin with a lemma about rejecting run trees for reactive safety properties.
\begin{lem}
\label{lem:helperlemma}
For a pruned nondeterministic parity tree automaton $\mathcal{A}$, representing an $I$/$O$-safe property, and a full tree $\langle T, \tau \rangle$ not in the language of $\mathcal{A}$, no run tree $\langle T_r, \tau_r \rangle$ for $\langle T, \tau \rangle$ has $t \in T_r$ for the violation starting node $t = t_0 \ldots t_n \in T$.
\end{lem}
\begin{proof}
We show the claim by assuming the converse and deriving a contradiction. In particular, we build a second full tree $\langle T', \tau' \rangle$ for which the path from the root to $t$ is the same as in $\langle T, \tau \rangle$, but that is accepted by $\mathcal{A}$ and thus contradicts the fact that $t$ is a violation starting node for $\mathcal{A}$. Without loss of generality, we assume that $t$ is a violation starting node that does not have a prefix which is also a violation starting node. 

We assume that $\mathcal{A} = (Q,I,O,\delta,q_0,\mathcal{F})$ for $Q = \{q_0, \ldots, q_m\}$. If $\mathcal{A}$ is pruned, then for every state $q_i$, there exists some full tree $\langle T^i, \tau^i\rangle$ that is in the language of $q_i$, with the corresponding run tree $\langle T^i_r, \tau^i_r \rangle$. 

In $\langle T', \tau' \rangle$, we replicate the path to the violation starting node of $\langle T, \tau \rangle$. We set $\tau'(\epsilon) = \tau(\epsilon)$ and $\tau'(t_0 \ldots t_i) = \tau(t_0 \ldots t_i)$ for all $i \in \{0, \ldots, n-1\}$. The corresponding accepting run tree $\langle T'_r, \tau'_r \rangle$ is also copied along this path, i.e., $\tau'_r(t_0 \ldots t_i x) = \tau(t_0 \ldots t_i x)$ for all $i \in \{0, \ldots, n-1\}$ and $x \in I$, and furthermore $\tau'_r(\epsilon) = \tau_r(\epsilon)$. This makes sure that $\langle T'_r, \tau'_r \rangle$ is a valid (and complete) prefix run tree for the parts of $\langle T, \tau \rangle$ defined so far. Note that the nodes of $\langle T_r, \tau_r \rangle$ referred to here are actually all well-defined as otherwise $t$ would have a prefix that is also a violation starting node. 

For the rest of $\langle T, \tau \rangle$, we copy the trees of the set $\{\langle T^0, \tau^0 \rangle, \ldots, \langle T^m, \tau^m \rangle\}$ declared above as sub-trees into $\langle T, \tau \rangle$ and set $\tau'(t_0 \ldots t_i x t') = \tau^k(t')$ for $k \in \{0, \ldots, m\}$ such that $q_k = \tau_r(t_0 \ldots t_i x)$ and all $i \in \{0, \ldots, n\}$, $x \in I$ and $t' \in I^*$. For the corresponding run tree $\langle T'_r, \tau'_r \rangle$, we do the same and set $\tau'_r(t_0 \ldots t_i x t') = \tau_r^k(t')$ for $k \in \{0, \ldots, n\}$ such that $q_k = \tau_r(t_0 \ldots t_i d)$ for all $i \in \{0, \ldots, n\}$, $x \in I$ and $t' \in I^*$. The resulting run tree is full and also accepting as all run trees in $\{\langle T^0_r, \tau^0_r \rangle, \ldots \langle T^m_r, \tau^m_r \rangle\}$, which form the suffix run trees in $\langle T'_r, \tau'_r \rangle$, are accepting.
\end{proof}
\begin{defi}
\label{WFunction}
Given a nondeterministic parity tree automaton $\mathcal{A} = (Q,I,O,\delta,q_0,\alpha)$, we define $\mathcal{W}(\mathcal{A}) = \mathcal{A}'$ for the deterministic safety word automaton $\mathcal{A}' = (Q',\Sigma,\delta',\{q_0\})$ for which $\Sigma = O \times I$, $Q' = 2^Q$ and for all $(x,y) \in \Sigma$ and $q,q' \in Q'$, we have $(q,(y,x),q') \in \delta'$ if and only if $q' = \{ \tilde q' \in Q \mid \exists \tilde q \in Q, f \in \delta(\tilde q,y): \tilde q \in q, f(x) = \tilde q'\}$. 
\end{defi}
\begin{theo}
\label{theo:safetypreserv}
The language of a pruned nondeterministic parity tree automaton
$\mathcal{A}$ is $I$/$O$-safe if and only if
$\mathcal{L}(\mathcal{T}_{I/O}(\mathcal{W}(\mathcal{A})))
= \mathcal{L}(\mathcal{A})$. Furthermore, if $\mathcal{L}(\mathcal{A})$ is $I$/$O$-safe, then $\mathcal{W}(\mathcal{A})$ is tight for
$\mathcal L(\mathcal A)$.
\end{theo}
\begin{proof}
$\Rightarrow$: 
Assume that some tree $\langle T, \tau \rangle$ is not accepted by $\mathcal{A}$. Since $\mathcal{A}$ represents an $I$/$O$-safety property, there must exist a violation starting node $t \in T$. As $\mathcal{A}$ is pruned, all run trees $\langle T_R, \tau_r \rangle$ thus need to have that $t \notin T_r$ (Lemma \ref{lem:helperlemma}). Since all rejected trees have this property, to check whether a tree is rejected, we thus only need to test whether any path in the tree necessarily leads to a corresponding finite maximal path in the run tree. By Definition \ref{WFunction}, $\mathcal{W}(\mathcal{A})$ rejects precisely these paths (due to the power-set construction involved) and is thus tight for $\mathcal{L}(\mathcal{A})$. By Definition \ref{TFunction}, $\mathcal{T}_{I/O}(\mathcal{W}(\mathcal{A}))$ rejects precisely the trees having such a path. Thus, the languages of $\mathcal{T}_{I/O}(\mathcal{W}(\mathcal{A}))$ and $\mathcal{A}$ are identical.

$\Leftarrow$: As the $\mathcal{T}_{I/O}$ function converts a safety word automaton into a deterministic safety tree automaton that accepts a tree if and only if all paths in the tree are accepted by the safety word automaton, any outcome of applying the $\mathcal{T}_{I/O}$ function is necessarily an $I$/$O$-safety property. As we assume that $\mathcal{L}(\mathcal{T}_{I/O}(\mathcal{W}(\mathcal{A})))
= \mathcal{L}(\mathcal{A})$, this means that $\mathcal{A}$ is also an $I$/$O$-safety property.
\end{proof}

We conclude the characterization of the regular reactive safety properties with the following theorem:
\begin{theo}
\label{thm:DSigmaAndDSTAreTheSame}
The set of regular $I$/$O$-safe properties coincides with the set of properties representable as deterministic safety tree automata with directions $I$ and labels $O$.
\end{theo}
\begin{proof}
$\Rightarrow$:
Assume that we have some regular $I$/$O$-safety property $\psi$ given. Since $\psi$ is regular, we can construct a nondeterministic parity tree automaton $\mathcal{A}$ from it, and by
Theorem~\ref{theo:safetypreserv}, a deterministic safety tree automaton with directions $I$ and labels $O$.

$\Leftarrow$: As a deterministic safety tree automaton accepts an $O$-labeled $I$-tree if its run tree is complete with respect to $I$, all trees that are not accepted by some deterministic safety tree automaton $\mathcal{A}$ have some finite maximal path in the run tree. Due to the determinism of $\mathcal{A}$, when taking the corresponding path in the rejected tree, copying this path into a different tree causes the new tree to be rejected by $\mathcal{A}$ as well.
\end{proof}

\section{Detecting Reactive Safety}

The goal of this section is to check if a given property, represented
as an automaton or an LTL formula, is a reactive safety property.  We
give separate constructions for tree and word properties.  The
algorithms of the first subsection analyze the languages of
nondeterministic and deterministic parity tree automata. The
algorithms of the second subsection analyze word languages that are
either given as LTL formulas or as nondeterministic Büchi automata.

\subsection{Reactive Safety for Tree Languages}

Our algorithm for nondeterministic parity tree automata is based on the observation that the language equality requirement in Theorem~\ref{theo:safetypreserv} can be weakened to language containment by the fact, shown in the following lemma, that the language of the tree automaton $\mathcal A$ is always contained in the language of $\mathcal{T}_{I/O}(\mathcal{W}(\mathcal{A}))$. We will show that this
condition can be checked in single-exponential time. Using Muller and
Schupp's complementation-by-dualization~\cite{Muller+Schupp/87/Dual},
we first obtain an automaton for the complement of
$\mathcal{L}(\mathcal{A})$. This language is then intersected with the
language of $\mathcal{T}_{I/O}(\mathcal{W}(\mathcal{A}))$, and the
emptiness of the resulting automaton is checked with a parity game.

\begin{lem}
\label{lem:ApplyingTandWLeadsToLanguageInclusion}
For a nondeterministic parity tree automaton $\mathcal{A}$, it holds that
$\mathcal{L}(\mathcal{A}) \subseteq \mathcal{L}(\mathcal{T}_{I/O}(\mathcal{W}(\mathcal{A})))$.
\end{lem}
\begin{proof}
By the construction of
$\mathcal{T}_{I/O}(\mathcal{W}(\mathcal{A}))$, we have that 
$\mathcal{L}(\mathcal{A}) \subseteq \mathcal{L}(\mathcal{T}_{I/O}(\mathcal{W}(\mathcal{A})))$,
because the
$\mathcal{W}$ function performs a power-set construction over
$\mathcal{A}$, so all missing paths in a run tree for
$\mathcal{T}_{I/O}(\mathcal{W}(\mathcal{A}))$ imply a missing
path in a run tree for $\mathcal{A}$. 
\end{proof}
Combining Theorem \ref{theo:safetypreserv} and Lemma \ref{lem:ApplyingTandWLeadsToLanguageInclusion}, we obtain that reactive safety can be characterized as language containment between $\mathcal{T}_{I/O}(\mathcal{W}(\mathcal{A}))$ and $\mathcal A$.
\begin{corol}
\label{corol:charac}
The language of a nondeterministic parity tree automaton
$\mathcal{A}$ is $I$/$O$-safe if and only if
$\mathcal{L}(\mathcal{T}_{I/O}(\mathcal{W}(\mathcal{A}))) \subseteq \mathcal{L}(\mathcal{A})$. 
\end{corol}
Using Corollary \ref{corol:charac}, we now devise an automata-theoretic algorithm for checking for reactive safety.
\begin{lem}
\label{lem:dual}
\cite{Muller+Schupp/87/Dual}
Given a nondeterministic parity tree automaton
$\mathcal{A}=(Q, I, O,\delta,q_0,\alpha)$ that runs on $O$-labeled $I$-trees,
the universal parity tree automaton
$\mathcal{U}=(Q,I, O,\delta,q_0,\alpha+1)$ accepts a tree $\langle I^*,\tau \rangle$ iff $\langle I^*, \tau \rangle$ is not accepted by $\mathcal{A}$.
\end{lem}

\begin{lem}
\label{lem:ndet}
\cite{DBLP:conf/lics/FinkbeinerS05,Muller+Schupp/95/Alternating}
Given a universal parity tree automaton $\mathcal{A}$ with $n$ states and $c$ colors, we can construct an equivalent nondeterministic parity tree automaton $\mathcal{N}$ with $n^{O(c \cdot n)}$ states and $O(c \cdot n)$ colors.
\end{lem}

\begin{theo}
\label{thm:NPTChecking}
Given a nondeterministic parity tree automaton $\mathcal{A} = (Q,I,O,\delta,q_0,\alpha)$, checking whether $\mathcal L(\mathcal{A})$ is $I$/$O$-safe (and obtaining a tight automaton for $\mathcal L(\mathcal A)$ in case of a positive result)  can be done in EXPTIME.
\end{theo}
\begin{proof}
As a first step, we identify and remove all states of $\mathcal{A}$
with an empty language. The emptiness check (by reduction to solving
parity games) can be done in time $n^{O(c)}$
\cite{DBLP:conf/stacs/Jurdzinski00}. Let the resulting automaton be called $\mathcal{A}'$.
By Corollary~\ref{corol:charac}, $\mathcal{A}'$ is
$I$/$O$-safe iff the language of
$\mathcal{T}_{I/O}(\mathcal{W}(\mathcal{A}'))$ is contained in the
language of $\mathcal{A}'$.  We check whether $\mathcal
L(\mathcal{T}_{I/O}(\mathcal{W}(\mathcal{A}'))) \cap
\overline{\mathcal L (\mathcal A')} = \emptyset$.  Applying
Lemma~\ref{lem:dual}, we translate $\mathcal A'$ into the universal
automaton $\mathcal U$ that recognizes the complement language.
$\mathcal U$ has the same size as $\mathcal A'$. Applying
Lemma~\ref{lem:ndet}, we obtain an equivalent nondeterministic
automaton $\mathcal{N}$ with $n^{O(c \cdot n)}$ states and $O(c \cdot
n)$ colors.  Computing the language intersection with the
deterministic automaton
$\mathcal{T}_{I/O}(\mathcal{W}(\mathcal{A'}))$, which has $2^{O(n)}$
states and a single color, we obtain the nondeterministic product
automaton $\mathcal P$ with $n^{O(c \cdot n)}$ states and $O(c \cdot
n)$ colors. 
The emptiness test of a nondeterministic parity tree 
automaton with $m$ states and $d$ colors can be done 
in $m^{O(d)}$ time \cite{DBLP:conf/stacs/Jurdzinski00}.
The overall time complexity is thus $n^{O(c^2 \cdot n^2)}$.
By Theorem~\ref{theo:safetypreserv}, $\mathcal{W}(\mathcal{A}')$ is tight for
$\mathcal L(\mathcal A')$ and thus also tight for $\mathcal L(\mathcal A)$.
\end{proof}

If the tree language is given as a deterministic automaton, we can
check whether the language is a reactive safety property with a
simpler construction, where we first prune states with empty languages
from the automaton and then search for a rejecting cycle in the
remaining state graph. This construction is analyzed in the following
theorem and will be used for the analysis of word languages in the
next subsection.

\begin{theo}
\label{thm:DPTChecking}
Given a deterministic parity tree automaton $\mathcal{A}$ over $I$/$O$ with $n$ states and $c$ colours, checking whether  $\mathcal L(\mathcal{A})$ is $I$/$O$-safe (and obtaining a tight automaton for $\mathcal L(\mathcal A)$ in case of a positive result) can be done in time $n^{O(c)}$.
\end{theo}
\begin{proof}
Again, as a first step, we identify and remove all states of $\mathcal{A}$
with an empty language. Let the resulting automaton be
called~$\mathcal{A}'$.  As a second step, we check if $\mathcal{A}'$
contains a rejecting cycle, which can be done in polynomial time
\cite{DBLP:conf/lata/Ehlers10}. $\mathcal{A}'$ contains a
rejecting cycle iff there exists an input tree that is rejected
and has a (unique) full run tree -- which is the case exactly if
$\mathcal L(\mathcal A')$, and hence $\mathcal L(\mathcal A)$, is not
safe.

To obtain the tight word automaton, we simply compute $\mathcal{W}(\mathcal{A}')$. For deterministic tree automata, the subset construction employed in Definition \ref{WFunction} does not increase the number of states in the automaton.
If $\mathcal{A}'$ does not contain any
rejecting loops, then $\mathcal L(\mathcal{T}_{I/O}(\mathcal{W}(\mathcal{A}'))= \mathcal L(\mathcal A')$, and, hence, $\mathcal{W}(\mathcal{A}')$ is tight for $\mathcal L(\mathcal A)$.
\end{proof}

\subsection{Reactive Safety for Word Languages}

We reduce the analysis of word languages, given as  LTL formulas
or as word automata, to the case of deterministic parity tree
automata solved in Theorem~\ref{thm:DPTChecking}. For this purpose,
we translate the given formula or automaton into a deterministic
parity automaton, which causes a doubly-exponential or single-exponential
blow-up, respectively, in the number of states.
  
\begin{theo}
\label{theo:LTL2EXPTIMEHardness}
Given a formula $\psi$ in linear-time temporal logic over the atomic propositions $\AP = \AP_I \cup \AP_O$, for $I = 2^{\AP_I}$ and $O = 2^{\AP_O}$, the problem of determining whether the set of $O$-labeled $I$-trees satisfying $\psi$ along all paths is $I$/$O$-safe (and obtaining a tight automaton in case of a positive result) is 2EXPTIME-complete.
\end{theo}
\begin{proof}
For the upper bound, we translate the LTL formula of size $n$ into a  deterministic parity word automaton $\mathcal A$ with
at most $2^{2^n \log n}$ states and $3(n+1)2^n$ colors \cite{VardiWilke-2007}. We then consider the tree automaton $\mathcal{T}_{I/O}(\mathcal{A})$, which has the same number of states and colors.
Applying Theorem~\ref{thm:DPTChecking}, we can thus check whether $\mathcal L(\mathcal A)$ is a reactive safety property and obtain the tight automaton in time
$2^{2^{O(n)}}$.

For the lower bound, we reduce the realizability problem of LTL, which is 2EXPTIME-complete \cite{DBLP:conf/popl/PnueliR89}, onto $I$/$O$-safety checking. Let $\psi$ be a specification over $\AP = \AP_I \cup \AP_O$ that is to be checked for realizability. We take $\psi' = \psi \wedge \mathsf{GF} a$ for some $a \notin \AP$. Then, $\psi'$ is realizable over $2^{\AP_I}$/$2^{\AP_O}$ if and only if $\psi$ is not $2^{\AP_I}$/$2^{\AP_O \cup \{a\}}$-safe:
\begin{itemize}
 \item If $\psi$ is realizable over $2^{\AP_I}$/$2^{\AP_O}$, then the $\mathsf{G F} a$ conjunct in $\psi'$ ensures that $\psi'$ is not $2^{\AP_I}$/$2^{\AP_O \cup \{a\}}$-safe.
 \item On the other hand, if $\psi$ is not realizable over $2^{\AP_I}$/$2^{\AP_O}$, so is $\psi'$ over $2^{\AP_I}$/$2^{\AP_O \cup \{a\}}$. As the empty tree property over $2^{\AP_I}$/$2^{\AP_O \cup \{a\}}$ has the property violation node $\epsilon$, $\psi$ is $2^{\AP_I}$/$2^{\AP_O \cup \{a\}}$-safe. 
\end{itemize}
\vspace{-5mm}
\end{proof}

\begin{theo}
Given a nondeterministic B\"uchi word automaton $\mathcal{A}$ over the alphabet $\Sigma = O \times I$, the problem of determining whether $\mathcal{A}$ is $I$/$O$-safe is EXPTIME-complete.
\end{theo}
\begin{proof}
For the upper bound, we translate the given nondeterministic B\"uchi
word automaton $\mathcal{A}$ into an equivalent deterministic parity
word automaton. If the B\"uchi automaton has $n$ states, the resulting
deterministic parity word automaton $\mathcal A'$ has at most $2^{O(n \log n)}$ states and
$2n+1$ colors \cite{DBLP:journals/lmcs/Piterman07,VardiWilke-2007}. Without changing the size of the automaton, we transform $\mathcal A'$ into the deterministic parity tree automaton $\mathcal{T}_{I/O}(\mathcal A')$  and apply
Theorem~\ref{thm:DPTChecking}: The check whether $\mathcal
L(\mathcal{A})$ is safe, and, in case of a positive result, the
construction of the tight automaton, thus takes at most
$2^{O(n^2 \log n)}$ time.

We obtain a matching lower bound from LTL realizability with a similar
reduction as in Theorem~\ref{theo:LTL2EXPTIMEHardness}. Since the
exponential-time hierarchy is strict, the translation from LTL
formulas to nondeterministic B\"uchi automata can be done with only an
exponential blow-up \cite{Vardi94reasoningabout}, and the LTL realizability problem is
2EXPTIME-complete, the realizability problem from nondeterministic
B\"uchi automata is EXPTIME-hard. We build an automaton for the LTL
formula $\mathsf{G F} a$. As taking the conjunction of two Büchi automata
results in only polynomial blow-up~\cite{Vardi1996}, the rest  of the
construction is analogous to the proof of Theorem~\ref{theo:LTL2EXPTIMEHardness}. 
\end{proof}

\subsection{The Coffee Machine Example}
\label{examplecontinued}

We finish this section with the coffee machine example from the
introduction. The specification is a conjunction $\psi_1 \wedge
\psi_2$ of two LTL formulas, $\psi_1 = \mathsf{G}(c \rightarrow
\mathsf{X}(f \vee \mathsf{F}b))$  (whenever the user
presses the coffee button, brewing must eventually start or a failure
must be signaled immediately) and $\psi_2 = \mathsf{G}(e
\rightarrow \mathsf{XG}( \neg b))$ (whenever the emergency shutdown button is pressed, brewing stops permanently), where
$c$ and $e$ are inputs and
$b$ and $f$ are outputs.

\tikzset{
	initial text=,
	every edge/.style={draw, thick, ->, auto},
	every state/.style={draw, thick, inner sep=0, minimum size=7mm}
}

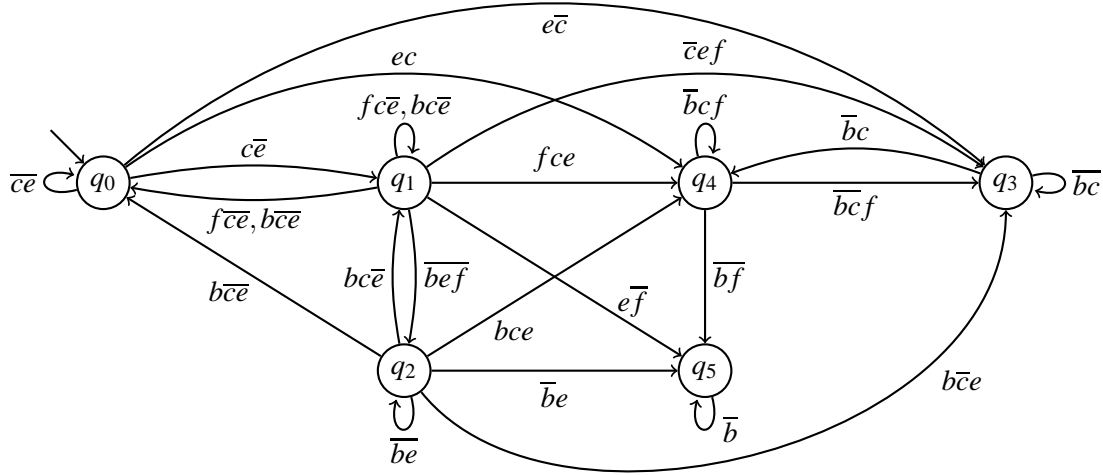
\begin{figure}
\centering \begin{tikzpicture}[scale=1.0]
		\useasboundingbox (-1.4,-4) rectangle (13.3,2.5);
        \node[state]               (l1) at (0,0) {$q_0$};
        \node[state]               (l2) at (4,0) {$q_1$};
        \node[state]               (l3) at (4,-2.5) {$q_2$};
        \node[state]               (l1b) at (12,0) {$q_3$};
        \node[state]               (l2b) at (8,0) {$q_4$};
        \node[state]               (l3b) at (8,-2.5) {$q_5$};
        \draw[thick,->] (-0.7,+0.7) -- (l1);
        \draw (l1) edge[bend left=10] node[above=-0.3mm] {$c \overline{e}$} (l2);
        \draw (l2) edge[bend left=10] node[below=-0.3mm] {$f \overline{ce}, b \overline{ce}$} (l1);
        \draw (l1) edge[loop left] node[left=-0.3mm] {$\overline{ce}$} (l1);
        \draw (l2) edge[loop above] node[above=-0.9mm] {$fc\overline{e}, bc\overline{e}$} (l2);
        \draw (l2) edge[bend left=10] node[right=-0.3mm] {$\overline{bef}$} (l3);
        \draw (l3) edge[bend left=10] node[left=-0.3mm] {$bc\overline{e}$} (l2);
        \draw (l3) edge node[below left=-0.8mm] {$b\overline{ce}$} (l1);
        \draw (l3) edge[loop below] node[below=-0.3mm] {$\overline{be}$} (l3);
        \draw (l1) edge[bend left=35] node[above=-0.3mm] {$ec$} (l2b);
        \draw (l1) edge[bend left=40] node[below] {$e\overline{c}$} (l1b);
        \draw (l2) edge node[above] {$fce$} (l2b);
        \draw (l3) edge node[below] {$\overline{b}e$} (l3b);
        \draw (l3) edge node[below right=-1.0mm, near start] {$bce$} (l2b);
        \draw (l2) edge node[near end, above right=-1.5mm] {$e \overline{f}$} (l3b);
        \draw (l2) edge[bend left=35] node[above=-0.3mm] {$\overline{c}ef$} (l1b);
        \draw[thick,->] (l3) .. controls (6,-5.0) and (12,-3.5) .. node [below right=-0.3mm, near end] {$b\overline{c}e$} (l1b);
        \draw (l2b) edge[loop above] node[above=-0.9mm] {$\overline{b}cf$} (l2b);
        \draw (l1b) edge[bend right=20] node[above=-0.3mm] {$\overline{b}c$} (l2b);
        \draw (l2b) edge node[below=-0.6mm] {$\overline{bc}f$} (l1b);
        \draw (l2b) edge node[right=-0.3mm] {$\overline{bf}$} (l3b);
        \draw (l3b) edge[loop below] node[right=0mm] {$\ \overline{b}$} (l3b);
        \draw (l1b) edge[loop right] node[right=-0.3mm] {$\overline{bc}$} (l1b);
        %
      \end{tikzpicture}
\caption{Deterministic parity word automaton $\mathcal A$ for the specification $\mathsf{G}(c \rightarrow \mathsf{X}(f \vee \mathsf{F}b)) \wedge \mathsf{G}(e \rightarrow \mathsf{XG}( \neg b))$. The states $q_2$ and $q_5$ have color $1$, the remaining states have color $2$. We use overlined atomic propositions to denote negated input or output bits. For example, the expression $f \overline{c}$ refers to all elements $x \in 2^{\AP_I \cup \AP_O}$ with $f \in x$ and $c \notin x$.}
\label{fig:ExampleDPW}
\end{figure}

The specification can be translated into the deterministic parity word
automaton $\mathcal{A}$ over the alphabet $2^{c,e,b,f}$ shown in Figure \ref{fig:ExampleDPW}. The
states $q_0$, $q_1$ and $q_2$ correspond to the case that the
emergency button (input $e$) has not been pressed yet. When the button
is pressed, the run of the automaton moves to the states $q_3$, $q_4$
and $q_5$, which mirror the behavior of $q_0$, $q_1$ and $q_2$, but
take into account that the emergency button has been pressed in the
past and the $b$ signal is therefore no longer allowed.

To check whether the language of $\mathcal{A}$ is a reactive safety property, we spread $\mathcal A$ to a tree automaton $\mathcal{A}' = (Q,I,O,\delta,q'_0,\mathcal{F})$ with the same set of states, and prune all states with empty language. In $\mathcal{A}'$, state $q_5$ has the empty language and is therefore removed. Note that this also removes all transitions $(q,y,f) \in \delta$ for which for some $x \in I$, $f(i)=q_5$. As a result, there are no transitions of the form $(q_1,\{b\},f)$ or $(q_1,\emptyset,f)$ anymore. Hence, state $q_2$ has become unreachable.

Since all remaining reachable states have color $2$, there are no infinite paths in the automaton on which the highest color occurring infinitely often is odd. Hence, the automata $\mathcal{A}$ and $\mathcal{A}'$ represent a reactive safety property. 

\section{Conclusion}

In this paper, we have extended the classic notion of linear-time
safety from closed systems, where all actions are under the system's
control to open reactive systems, where the behavior is characterized
by the interplay of uncontrolled environment inputs and controlled
system outputs. Reactive safety is a larger class of properties than
standard linear-time safety; at the same time, the algorithmic
advantages are retained, because it is still possible to translate any
(regular) reactive safety property into a safety word automaton, which
can be used, for example, as a runtime monitor. In fact, reactive
safety is the maximal set of properties whose satisfaction can be
checked by testing all computation paths against a linear-time safety
property. It is conceivable, however, to further extend the class of
safety properties if other systems aspects, beyond the inputs and
outputs, are taken into consideration. A promising candidate is
incomplete information: specifications are sometimes concerned with
atomic propositions that can neither be read nor written to by the
system.  Such an extension would classify an even larger set of
properties as safety. Extending the algorithms of this paper to this
case is straightforward using standard automata-theoretic techniques
for synthesis under incomplete information~\cite{KV97c}.

\bibliographystyle{eptcs}
\bibliography{bib}
\end{document}